\documentclass[reprint, amsmath, amssymb, aps, pre]{revtex4-2}

\usepackage[utf8]{inputenc}
\usepackage[T1]{fontenc}
\usepackage{graphicx}
\usepackage{xcolor}
\usepackage{mathtools}
\usepackage{bm}
\usepackage{hyperref}
\usepackage{ulem}
\newcommand{\dd}{\mathrm{d}}
\newcommand{\Dlow}{D_{\mathrm{low}}}
\newcommand{\Dhigh}{D_{\mathrm{high}}}
\newcommand{\Dbar}{\bar D}
\newcommand{\E}{\mathbb{E}}

\begin{document}

\title{Switching diffusivity selects Pareto tail exponent in random growth with redistribution}

\author{Maxence Arutkin}
\affiliation{School of Chemistry, Tel Aviv University, Tel Aviv, Israel}
\affiliation{Physics Division, National Center for Theoretical Sciences, National Taiwan University, Taipei 106319, Taiwan}
\author{Alexandre Vall\'ee}
\affiliation{Department of Epidemiology and Public Health, Foch Hospital, Suresnes, France}
\date{\today}

\begin{abstract}
Random multiplicative growth with redistribution generates stationary Pareto wealth tails in the Bouchaud--M\'ezard model, but assumes a fixed multiplicative noise intensity. This is restrictive for physical and financial growth processes, where volatility (diffusivity) is often fluctuating. We replace the constant noise intensity by a switching diffusivity and ask how these fluctuations select the Pareto stationary tail. For a geometric Brownian motion with switching diffusivity, the long-time Gaussian limit holds when the redraw law has finite mean and variance. The asymptotic variance retains a contribution from diffusivity persistence. With redistribution and a general redraw law, the stationary large-wealth problem is characterized by a spectral condition for admissible algebraic modes. For a two-state diffusivity, an exact tail analysis gives a Pareto exponent interpolating between the high-diffusivity slow-refresh limit and the mean-diffusivity fast-refresh Bouchaud--M\'ezard limit. 
\end{abstract}

\maketitle

\section{Introduction}

Stationary heavy-tailed distributions are a hallmark of socioeconomic systems, from individual wealth \cite{druagulescu2001exponential,yakovenko2009colloquium,charpentier2022pareto} and firm sizes to city populations \cite{gabaix1999zipf,gabaix2009power}. A physical mechanism generating such distributions is random multiplicative growth stabilized by an additive restoring term~\cite{sornette1997convergent,zanette2020fat}. Its mean-field realization, the Bouchaud--M\'ezard exchange economy~\cite{bouchaud2000wealth,biham1998generic,levy1996power,chatterjee2007kinetic,greenberg2024twenty,boghosian2014kinetics}, extensively studied on networks \cite{garlaschelli2008effects,ichinomiya2012bouchaud,ichinomiya2013power}, and recently extended to heterogeneous mean growth rates~\cite{bernard2026mean}, balances pairwise redistribution against spontaneous multiplicative fluctuations. It produces a stationary inverse-gamma density with Pareto exponent $\alpha = 1 + J/D$, set by two parameters: the redistribution rate $J$ and the multiplicative noise intensity $D$ that we call here diffusivity, consistent with the general Kesten mechanism~\cite{kesten1973random}. This result relies on the implicit assumption that the diffusivity $D$ is a constant. More generally, the role of heterogeneity in random-growth models extends beyond the stationary distribution. Gabaix et al.~\cite{gabaix2016dynamics} showed that heterogeneity in return statistics can be essential for reproducing the observed speed of inequality dynamics. Whereas their analysis concerns the temporal evolution of inequality, we address here a complementary stationary question: how temporal persistence in the variance of multiplicative growth modifies the Pareto tail even at fixed mean diffusivity.

From single-molecule tracers in heterogeneous or active media~\cite{wang2012brownian, chubynsky2014diffusing, chechkin2017brownian, tyagi2017non, sposini2018random, jain2016diffusion, lanoiselee2018diffusion,cherstvy2016anomalous} to returns in financial markets~\cite{gopikrishnan1999scaling,cont2001empirical}, the instantaneous variance governing fluctuations is itself a random variable~\cite{heston1993closed}, with persistence times comparable to those of the observable process~\cite{cont2001empirical,ding1993long}. The diffusing-diffusivity framework and its reformulation as a doubly stochastic continuous-time random walk (DSCTRW) \cite{arutkin2024doubly} have established that such latent diffusivity generates pronounced non-Gaussian transients in additive Brownian motion while preserving Gaussian long-time statistics through disorder self-averaging. We ask whether this self-averaging survives the multiplicative nonlinearity and the redistributive interaction.

For a single multiplicative process, the logarithmic process reduces exactly to a subordinated Brownian motion controlled by the integrated diffusivity $\Lambda_t = \int_0^t D_s\,\dd s$. The switching diffusivity then produces non-Gaussian statistics at short times, while the centered logarithmic process self-averages and becomes Gaussian at long times when the diffusivity has finite mean and variance. For an arbitrary redraw law $f_D$, we derive the Fourier--Laplace propagator in renewal form and characterize the resulting short and long-time regimes. For the bounded two-state law used in the explicit results below, the short-time distribution is a Gaussian mixture, while at long times the integrated diffusivity self-averages and the centered logarithmic process becomes Gaussian. The exponential redraw law, for which the propagator can be evaluated explicitly but annealed multiplicative moments can diverge at finite times, is treated separately in the Supplemental Material. 

For an ensemble of agents coupled by mean-field redistribution, the latent diffusivity trajectory persists long enough to modulate the competition between multiplicative spreading and restoring exchange, and the stationary large-wealth problem retains information about the diffusivity statistics beyond their mean. For an arbitrary redraw law $f_D$, we derive a spectral condition for admissible algebraic stationary modes. At the macroscopic level, Markov-switching growth models have established that persistence can affect Pareto exponents~\cite{beare2022determination}. Here we determine how persistence in the multiplicative noise intensity changes the stationary wealth tail under Bouchaud--M\'ezard redistribution. For two-state diffusivity, keeping the redistribution rate and the entire diffusivity distribution fixed, we find that the Pareto exponent lies strictly below the mean-diffusivity prediction $1+J/\bar D$ at every finite refresh rate. Thus, continuous redistribution does not eliminate the effect of diffusivity persistence on the stationary tail: replacing $D_t$ by its mean systematically underestimates the stationary tail probability at sufficiently large wealth. We determine the exponent throughout the crossover from $1+J/\Dhigh$ at slow refresh to $1+J/\bar D$ at fast refresh. In both limits, we also derive the complete normalized stationary distributions: a mixture of the two fixed-diffusivity Bouchaud--M\'ezard distributions at slow refresh, and the single Bouchaud--M\'ezard distribution with diffusivity $\bar D$ at fast refresh.

\section{Multiplicative Growth with Switching Diffusivity}
\label{sec:bare}
\subsection{Definition of the model}

We first characterize the dynamics of a single agent's wealth $z_t > 0$ in the absence of redistribution. The multiplicative model is given by the stochastic differential equation
\begin{equation}\label{eq:part1-model}
    \dd z_t = m z_t\,\dd t + \sqrt{2D_t}\,z_t\,\dd B_t\,,
\end{equation}
where $m$ is a constant drift, $B_t$ is a standard Brownian motion, and the diffusivity $D_t \ge 0$ is a piecewise-constant stochastic process defined by $\{T_n\}_{n\ge 1}$  the jump times of an independent Poisson process of rate $\mu_r$, with $T_0=0$. The initial diffusivity $D_1\sim f_D$ is independent of $B_t$. On each interval $[T_n,T_{n+1})$, the diffusivity is constant. At each refresh event, a new value is drawn independently from a prescribed redraw law $f_D(D)$. We operate throughout in the It\^o convention. For the multiplicative amplitude $\sqrt{2D_t}\,z_t$, the Stratonovich formulation is not equivalent and would differ by the drift correction $D_t z_t\,\dd t$. All results below are therefore stated for the It\^o dynamics. 

To linearize Eq.~\eqref{eq:part1-model}, we apply It\^o's formula to the logarithmic variable $y_t = \ln z_t$. With $g(z) = \ln z$, $g'(z) = 1/z$, and $g''(z) = -1/z^2$, the quadratic variation term is $(\dd z_t)^2 = 2D_t z_t^2\,\dd t$. Substitution yields the additive equation
\begin{equation}\label{eq:Y-sde-generic}
    \dd y_t = (m-D_t)\,\dd t + \sqrt{2D_t}\,\dd B_t\,.
\end{equation}
Integrating from $0$ to $t$, we introduce the integrated diffusivity
\begin{equation}\label{eq:Lambda-def}
    \Lambda_t = \int_0^t D_s\,\dd s\,,
\end{equation}
yielding the pathwise representation
\begin{equation}\label{eq:Y-representation}
    y_t = y_0 + m t - \Lambda_t + \int_0^t \sqrt{2D_s}\,\dd B_s\,.
\end{equation}
Conditional on the entire realization of the volatility path $\{D_t\}$, $\Lambda_t$ is deterministic. By It\^o isometry, the stochastic integral $\int_0^t \sqrt{2D_s}\,\dd B_s$ is Gaussian with mean zero and variance $2\Lambda_t$. Therefore,
\begin{equation}\label{eq:Y-cond-Gaussian}
    p(y_t\mid \{D_t\}) \sim \mathcal N\left(y_0+m t-\Lambda_t,\;2\Lambda_t\right)\,.
\end{equation}
This establishes the direct multiplicative analog to the diffusing diffusivity/DSCTRW framework: the law of the logarithmic wealth $y_t$ is obtained by mixing Gaussian kernels whose mean and variance are strictly controlled by $\Lambda_t$.  For the explicit results below, we use the bounded two-state redraw law
\begin{equation}  f_D(D)=\beta\,\delta(D-\Dlow)+(1-\beta)\,\delta(D-\Dhigh),  \label{eq:two_state_nu} \end{equation} 
with $0<\Dlow<\Dhigh$ and $0<\beta<1$.

\begin{figure}[t]
    \centering
    \includegraphics[width=\columnwidth]{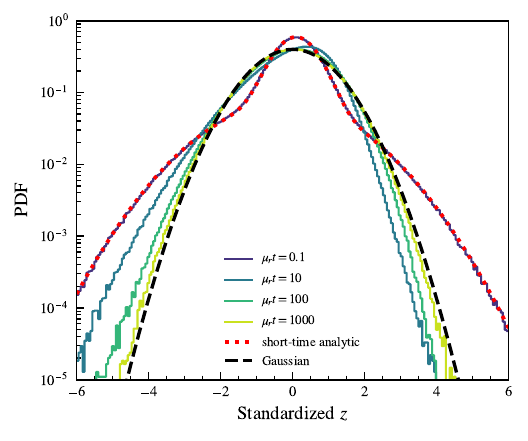}
    \caption{Standardized centered log-return distribution for the two-state switching-diffusivity model defined in Eq.~\eqref{eq:two_state_nu}. The plotted variable is $z=[r-\mathbb{E}(r)]/\sqrt{\mathrm{Var}(r)}$, with $r=y_t-y_0-mt$. Parameters are $\Dlow=0.12$, $\Dhigh=1.10$, and $\beta=0.72$. Curves are shown for $\mu_r t=0.1,\,10,\,100,$ and $1000$, with $\mu_r=1$ ($t=0.1,\,10,\,100,$ and $1000$), and are obtained by exact fixed-time sampling of $\Lambda_t$ without time discretizations. For $\mu_r t\ll1$, the law retains the non-Gaussian shape of a frozen-diffusivity Gaussian mixture. The dotted curve is the analytical frozen-diffusivity mixture at $\mu_r t=0.1$; the dashed curve is the asymptotic unit Gaussian. For $\mu_r t\gg1$, centering and variance normalization collapse the distribution onto the unit Gaussian, showing self-averaging of the integrated diffusivity.
}
    \label{fig:gaussian_crossover}
\end{figure}

\subsection{Exact propagator and asymptotic regimes}

We first derive the propagator for an arbitrary redraw law $f_D(D)$. The contribution of trajectories with exactly $n$ refreshes in $[0,t]$ is obtained by partitioning the interval into $n+1$ ordered plateaux of lengths $\tau_1,\dots,\tau_{n+1}$ with independent diffusivities $D_1,\dots,D_{n+1}$. The corresponding measure for $(\tau_1,\dots,\tau_{n+1},D_1,\dots,D_{n+1})$ is
\begin{equation}\label{eq:joint-full}
    e^{-\mu_r t}\,\mu_r^n\,\delta\!\Bigl(t-\sum_{j=1}^{n+1}\tau_j\Bigr)\prod_{j=1}^{n+1}\mathbf{1}_{\{\tau_j>0\}}\,f_D(D_j)\,.
\end{equation}
The log-increment on this configuration is
\begin{equation}
    y_t-y_0 = m t - \sum_{j=1}^{n+1}D_j\tau_j + \sum_{j=1}^{n+1}\sqrt{2D_j\tau_j}\,\xi_j\,,
\end{equation}
with $\xi_j$ i.i.d.\ standard Gaussian. The characteristic function $\Psi(k,t) = \E[e^{ik(y_t-y_0)}]$ is obtained by integrating $e^{ik(y_t-y_0)}$ against \eqref{eq:joint-full} and the Gaussian law of the $\xi_j$:
\begin{equation}\label{eq:Psi-joint}
\begin{split}
    \Psi(k,t) = \sum_{n=0}^{\infty}\mu_r^n\,e^{-\mu_r t}\!\int\!\dd\boldsymbol{\tau}\,\dd\boldsymbol{D}\,\delta\!\Bigl(t-\sum_{j=1}^{n+1}\tau_j\Bigr) \\
    \times \prod_{j=1}^{n+1}\mathbf{1}_{\{\tau_j>0\}}\,f_D(D_j)\,\E_{\xi_j}\!\bigl[e^{ik(m\tau_j - D_j\tau_j + \sqrt{2D_j\tau_j}\,\xi_j)}\bigr]\,.
\end{split}
\end{equation}

Laplace-transforming \eqref{eq:Psi-joint} in $t$, the delta factorizes $e^{-st}$ as $\prod_j e^{-s\tau_j}$, which combines with $e^{-\mu_r t} = \prod_j e^{-\mu_r\tau_j}$ to make the $n+1$ plateaux integrals fully separable. Each plateau contributes the same factor $\tilde\phi(k,s)$, and summing the geometric series in $n$ yields
\begin{equation}\label{eq:G-exp-Laplace}
    \widetilde\Psi(k,s)=\frac{\tilde\phi(k,s)}{1-\mu_r\tilde\phi(k,s)}\,.
\end{equation}
This has the renewal structure of a Montroll--Weiss propagator~\cite{montroll1965random}, with plateaux replacing waiting-time jumps. A closely related renewal structure was recently obtained by Guéneau et al.~\cite{gueneau2025large} for switching diffusion, where the diffusion coefficient is redrawn from a prescribed distribution at a constant rate. They showed that the far tails are dominated by rare trajectories that remain at the largest diffusivity without switching. In the present multiplicative setting, the same renewal structure implies that extreme wealth-growth trajectories are likewise controlled by prolonged high-diffusivity episodes. The per-plateau factor is itself a Gaussian average over $\xi$, an average over $D$ against $f_D$, and a Laplace integral over $\tau$:
\begin{equation}\label{eq:phi-tilde-def}
    \tilde\phi(k,s) = \int_0^\infty\!\dd\tau\,e^{-(s+\mu_r-ikm)\tau}\!\int_0^\infty\!\dd D\,f_D(D)\,e^{-(k^2+ik)D\tau}\,.
\end{equation}
Performing the integral over the plateau duration gives the equivalent representation
\begin{equation}\label{eq:phi-tilde-integrated}
\tilde\phi(k,s)=\int_0^\infty\frac{f_D(D)}{s+\mu_r-ikm+(k^2+ik)D}\,\dd D .
\end{equation}

For $t\ll\mu_r^{-1}$, the diffusivity is effectively frozen over the observation time. Writing $r=y_t-y_0-mt$, the distribution is therefore the Gaussian mixture
\begin{align}
p_{\mathrm{short}}(r,t)
&=
\frac{\beta}{\sqrt{4\pi \Dlow t}}
\exp\!\left[-\frac{(r+\Dlow t)^2}{4\Dlow t}\right]
\nonumber \\
&
\quad +
\frac{1-\beta}{\sqrt{4\pi \Dhigh t}}
\exp\!\left[-\frac{(r+\Dhigh t)^2}{4\Dhigh t}\right].
\end{align}
Thus the non-Gaussian short-time law is an ensemble mixture of Gaussian populations with different diffusivities.

At long times, the result is more general. For any redraw law with finite mean $\bar D$ and variance $\sigma_D^2$, Poisson renewal gives
$\mathrm{Cov}(D_s,D_{s+u})=\sigma_D^2e^{-\mu_r u}$. Consequently, the centered logarithmic process converges to
\begin{equation}
\frac{y_t-y_0-(m-\bar D)t}{\sqrt{t}}\sim\mathcal N\!\left(0,2\bar D+\frac{2\sigma_D^2}{\mu_r}\right).
\end{equation}
Figure~\ref{fig:gaussian_crossover} illustrates this crossover for the two-state model, from the frozen Gaussian mixture at short times to the Gaussian regime produced by self-averaging of the integrated diffusivity at long times. The distributions are sampled directly at fixed time through $\Lambda_t$, as detailed in the Supplemental Material.

The exponential redraw law is treated separately in the Supplemental Material, where Eq.~\eqref{eq:phi-tilde-def} can be evaluated explicitly and the short and long-time regimes are obtained in closed form. Its unbounded support also produces a finite-time divergence of the annealed multiplicative moments. This motivates the bounded two-state law used below. For the isolated multiplicative process, diffusivity persistence therefore controls finite-time fluctuations but self-averages in the logarithmic variable at long times. We now turn to the interacting problem, where redistribution generates a stationary relative-wealth distribution and diffusivity persistence can affect its Pareto tail.

\section{Interacting Growth with Redistribution}
\label{sec:interacting}

Given the single-agent dynamics of Sec.\ref{sec:bare}, we now turn to the interacting problem. The single-agent analysis describes how switching diffusivity affects the logarithmic process, but it cannot by itself address stationary Pareto tails. Such tails arise in the Bouchaud--M\'ezard model because multiplicative growth is balanced by redistribution between agents. 

We couple $N$ agents subject to doubly stochastic growth via a redistribution matrix $J_{ij}$. The wealth of agent $i$ obeys
\begin{equation}
\dd z_i = m_i z_i\,\dd t + \sum_{j\neq i}\bigl(J_{ij}z_j-J_{ji}z_i\bigr)\,\dd t + \sqrt{2D_i(t)}\,z_i\,\dd B_i.
\label{eq:full_interacting}
\end{equation}

We impose the mean-field scaling $J_{ij}=J/N$ for $i\neq j$, with $J>0$ the global redistribution rate, and set the deterministic drift $m_i=m$ for all $i$. The switching diffusivities $D_i(t)$ are refreshed independently across agents.

To isolate relative fluctuations, we define the mean wealth $\bar z = N^{-1}\sum_{k=1}^N z_k$. Averaging Eq.~\eqref{eq:full_interacting} over the population gives:
\begin{equation}
\dd\bar z = m\bar z\,\dd t + \frac{J}{N}\sum_{i=1}^N(\bar z-z_i)\,\dd t + \frac{1}{N}\sum_{i=1}^N \sqrt{2D_i(t)}\,z_i\,\dd B_i\,.
\end{equation}
The last term is the residual noise in the population average. Since it is a sum of independent individual shocks divided by $N$, it vanishes in the large-population limit provided no finite group of agents carries a finite fraction of the total wealth. We therefore take the mean wealth to follow deterministic macroscopic growth, $\dd\bar z \simeq m\bar z\,\dd t$. This reduction is self-consistent in the regime selected below. The stationary Pareto tail has $\alpha_c>1$, so the largest relative wealth scales as $x_{\max}\sim N^{1/\alpha_c}$. Hence $x_{\max}/N\to 0$ as $N\to\infty$, and no finite group of agents carries a finite fraction of the total wealth.

Using this large-population approximation, we define $x_i = z_i/\bar z$. Applying It\^o's formula to the ratio, it removes the common growth term $m$. Dropping the agent index, the effective single-agent dynamics under redistribution becomes:
\begin{equation}\label{eq:x-eff}
\dd x = J(1-x)\,\dd t + \sqrt{2D_t}\,x\,\dd B_t\,.
\end{equation}
Equation~\eqref{eq:x-eff} is the central stochastic equation for the relative wealth with switching diffusivity. The restoring drift $J(1-x)$ now competes directly with the multiplicative noise modulated by $D_t$.

\subsection{Coupled stationary Fokker-Planck system}

We first keep the redraw law $f_D(D)$ arbitrary. The joint process $(x_t, D_t)$ is Markov while $x_t$ alone is not, so we write the forward equation for the joint density $p(x,D,t)$.  
It reads
\begin{equation}
\partial_t p=-\partial_x\!\bigl[J(1-x)p\bigr]+\partial_x^2\!\bigl[D x^2p\bigr]+\mu_r\bigl[f_D(D)P(x,t)-p\bigr].
\label{eq:joint_fp_xD}
\end{equation}
The large-$x$ stationary problem can also be characterized without specifying $f_D$. Seeking an algebraic mode $p(x,D)\sim a_\alpha(D)x^{-1-\alpha}$ gives the spectral condition
\begin{equation}
\mu_r\int_0^\infty\frac{f_D(D)}{\mu_r+\alpha J-\alpha(\alpha-1)D}\,\dd D = 1.
\end{equation}
This condition is directly related to the renewal kernel derived in Sec.~\ref{sec:bare}. At large $x$, the constant part of the redistribution drift is negligible, so that $J(1-x)\simeq-Jx$. The asymptotic dynamics therefore coincide with the isolated multiplicative process with drift $m=-J$. Evaluating Eq.~\eqref{eq:phi-tilde-integrated} at $k=-i\alpha$ and $s=0$ gives
\begin{equation}\label{eq:phi-tail}
\tilde\phi(-i\alpha,0)=\int_0^\infty\frac{f_D(D)}{\mu_r+\alpha J-\alpha(\alpha-1)D}\,\dd D .
\end{equation}
The stationary spectral condition can therefore be written equivalently as
\begin{equation}\label{eq:renewal-tail-condition}
1-\mu_r\tilde\phi(-i\alpha,0)=0 .
\end{equation}
Thus, the same renewal denominator that governs the propagator of the isolated multiplicative process also selects the algebraic stationary modes in the presence of redistribution. Under the logarithmic transformation $y=\ln x$, an algebraic stationary mode $P(x)\sim x^{-1-\alpha}$ corresponds exactly to an exponential mode $p_y(y)\sim e^{-\alpha y}$. The condition for such a mode therefore coincides with a zero of the Montroll--Weiss denominator at $k=-i\alpha$ and $s=0$. In the large-deviation picture of Guéneau et al.~\cite{gueneau2025large}, this means that the rare growth trajectories feeding the high-wealth tail must compensate the contraction produced by redistribution, for a Pareto mode of exponent $\alpha$, this contraction occurs at rate $\alpha J$. Among the algebraic modes present in the stationary distribution, the smallest $\alpha>1$ is the slowest decaying and therefore controls the far tail. A formal solution corresponds to an admissible stationary tail mode only when the associated amplitude is non-negative and integrable over the support of $f_D$. The derivation and admissibility conditions are given in the Supplemental Material. We now specialize to the two-state law of Eq.~\eqref{eq:two_state_nu}, for which the physical tail can be selected explicitly. The marginal density of $x_t$ is $P(x,t)=\int p(x,D,t)\,\dd D$. For the two-state law, at stationarity $P(x)=P_\ell(x)+P_h(x)$. Projecting the joint forward equation onto $D=\Dlow$ and $D=\Dhigh$ gives the stationary balance equations:
\begin{align}
0 =& -\partial_x\!\bigl[J(1-x)P_\ell\bigr] + \Dlow\,\partial_x^2\!\bigl[x^2P_\ell\bigr] \nonumber \\
& +\mu_r\bigl[\beta P(x)-P_\ell\bigr], \label{eq:stat_low} \\
0 =& -\partial_x\!\bigl[J(1-x)P_h\bigr] + \Dhigh\,\partial_x^2\!\bigl[x^2P_h\bigr] \nonumber \\
& +\mu_r\bigl[(1-\beta)P(x)-P_h\bigr] . \label{eq:stat_high}
\end{align}
In the low-diffusivity channel, probability leaves at rate $\mu_r P_\ell$, while all agents at position $x$, irrespective of their previous diffusivity state, are redrawn into the low channel with probability $\beta$, giving an incoming rate $\mu_r\beta P(x)$. The high-diffusivity channel is analogous.

\subsection{Stationary tail and Pareto exponent}

For large $x$, Eqs.~\eqref{eq:stat_low}-\eqref{eq:stat_high} reduce to a system whose drift coefficient is linear in $x$, diffusion coefficient quadratic in $x$, and refresh terms constant. Under the change of variable $u = \ln x$, this system becomes linear with constant coefficients, whose solutions are exponentials $e^{\lambda u} = x^\lambda$. We therefore look for algebraic tail modes $P_i(x) \sim a_i x^{-1-\alpha}$.

Suppose $P_\ell \sim a_\ell x^{-1-\alpha_\ell}$ and $P_h \sim a_h x^{-1-\alpha_h}$ with $\alpha_\ell \neq \alpha_h$. Without loss of generality, $\alpha_\ell < \alpha_h$, so $P_\ell$ decays more slowly than $P_h$ at large $x$. In the equation for $P_h$, the refresh term $\mu_r(1-\beta)P_\ell$ then decays as $x^{-1-\alpha_\ell}$, while every other term decays as $x^{-1-\alpha_h}$, that is faster. This slower term has nothing to balance it, so the equation forces $P_h$ to contain a contribution decaying as $x^{-1-\alpha_\ell}$. This contradicts the assumption that $P_h$ decays as $x^{-1-\alpha_h}$. Both channels must therefore share a common exponent.

We insert $P_i(x) = a_i x^{-1-\alpha}$ in the large-$x$ asymptotic regime. Using $\partial_x[xf] = -\alpha x^{-1-\alpha}$ and $\partial_x^2[x^2 f] = \alpha(\alpha-1) x^{-1-\alpha}$ for $f(x) = x^{-1-\alpha}$, the drift and diffusion in channel $i$ give $-F_i(\alpha) a_i x^{-1-\alpha}$, with
\begin{equation}
F_i(\alpha) = \alpha J - \alpha(\alpha-1) D_i.
\end{equation}

Substituting in the two equations and dividing by $x^{-1-\alpha}$ gives a linear system for the amplitudes,
\begin{equation}
M_\alpha \begin{pmatrix} a_\ell \\ a_h \end{pmatrix} = 0,
\end{equation}
with
\begin{equation}
M_\alpha = \begin{pmatrix} F_\ell(\alpha) + \mu_r(1-\beta) & -\mu_r\beta \\ -\mu_r(1-\beta) & F_h(\alpha) + \mu_r\beta \end{pmatrix}.
\end{equation}
A nonzero tail requires $\det M_\alpha = 0$. Expanding the determinant, the $\mu_r^2$ terms cancel. Using $\bar D = \beta D_\ell + (1-\beta) D_h$,
\begin{equation}
\label{eq:Delta-explicit}
\Delta_\alpha = F_\ell(\alpha) F_h(\alpha) + \mu_r[\alpha J - \alpha(\alpha-1) \bar D] = 0.
\end{equation}
The smallest root $\alpha > 1$ of $\Delta_\alpha = 0$ is the tail exponent $\alpha_c$, in line with spectral characterizations of Pareto exponents in Markov multiplicative processes~\cite{Kontoyiannis,beare2022determination}.

In the fast-refresh limit $\mu_r\to\infty$, the diffusivity changes on a time scale much shorter than the relaxation time $J^{-1}$ of the relative wealth. During the time over which $x$ changes significantly, each agent has already visited the two diffusivity states many times. The multiplicative noise therefore sees only the averaged intensity
\begin{equation}
\Dbar=\beta\Dlow+(1-\beta)\Dhigh .
\end{equation}
More explicitly, introducing $Q=P_\ell-\beta P$ in the stationary equations gives  at leading order for fast refresh
\begin{equation}
Q\simeq-\frac{\beta(1-\beta)(\Dhigh-\Dlow)}{\mu_r}\,
\partial_x^2(x^2P),
\end{equation}
so that $Q\to0$ as $\mu_r\to\infty$. The marginal equation then reduces to the Bouchaud--M\'ezard equation with effective diffusivity $\Dbar$. Its normalized stationary distribution is
\begin{equation}
P_{\mathrm{fast}}(x)=\frac{(J/\Dbar)^{1+J/\Dbar}}{\Gamma(1+J/\Dbar)}x^{-2-J/\Dbar}e^{-J/(\Dbar x)} .
\end{equation}
Its Pareto exponent is therefore
\begin{equation}\label{eq:q-fast}
\alpha_c^{\mathrm{fast}}=1+\frac{J}{\Dbar}.
\end{equation}
The finite-refresh correction obtained from the same expansion is given in the Supplemental Material.

In the slow-refresh regime, the diffusivity does not self-average on the relaxation time $J^{-1}$ of the relative wealth. Taking the stationary limit first and then $\mu_r\to0$, each diffusivity channel approaches its fixed-diffusivity Bouchaud--M\'ezard stationary state. The complete limiting distribution is therefore
\begin{align}
P_{\mathrm{slow}}(x)=&\beta\frac{(J/\Dlow)^{1+J/\Dlow}}{\Gamma(1+J/\Dlow)}x^{-2-J/\Dlow}e^{-J/(\Dlow x)}\nonumber\\
+&(1-\beta)\frac{(J/\Dhigh)^{1+J/\Dhigh}}{\Gamma(1+J/\Dhigh)}x^{-2-J/\Dhigh}e^{-J/(\Dhigh x)} .
\end{align}
A larger diffusivity does not increase the mean relative wealth: both states are subject to the same restoring drift $J(1-x)$. What changes is the strength of the multiplicative fluctuations. A larger $D$ makes large relative-wealth excursions more likely, and therefore gives a heavier Pareto tail. The leading finite-refresh correction is derived in the Supplemental Material.

Taking the slow-refresh limit directly in Eq.~\eqref{eq:Delta-explicit} gives
\begin{equation}
\Delta_\alpha \to F_\ell(\alpha) F_h(\alpha),
\end{equation}
whose admissible roots above $1$ are $\alpha_\ell = 1 + J/\Dlow$ and $\alpha_h = 1 + J/\Dhigh$. Of these two candidate exponents, the one with the smaller exponent decays more slowly and therefore dominates at large $x$. Since $\Dhigh > \Dlow$, this is
\begin{equation}\label{eq:q-slow}
\alpha_c^{\mathrm{slow}} = 1 + \frac{J}{\Dhigh}.
\end{equation}

\begin{figure}[t]
    \centering
    \includegraphics[width=\columnwidth]{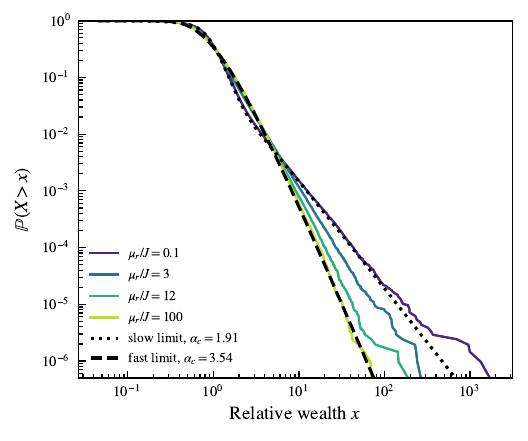}
    \caption{
Stationary complementary cumulative distribution $\mathbb{P}(X>x)$ for the effective relative-wealth dynamics Eq.~\eqref{eq:x-eff} with two-state switching diffusivity. Parameters are $J=1$, $\Dlow=0.12$, $\Dhigh=1.10$, $\beta=0.72$, and $\Dbar=0.3944$, giving $\alpha_c^{\mathrm{slow}}\simeq1.91$ and $\alpha_c^{\mathrm{fast}}\simeq3.54$. The colored curves are stationary numerical samples for $\mu_r/J=0.1,\,3,\,12,$ and $100$. The dashed and dotted black curves are the analytical fast- and slow-refresh limiting CCDFs, respectively. As refresh becomes faster, the distribution approaches the fast-refresh limit; as refresh slows down, it approaches the frozen-mixture limit.
}
    \label{fig:tail_crossover}
\end{figure}

\begin{figure}[t]
    \centering
    \includegraphics[width=\columnwidth]{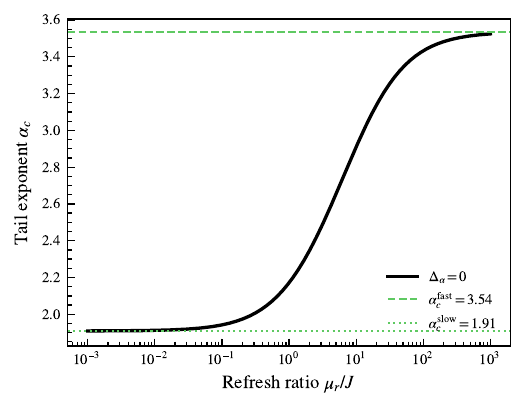}
    \caption{Stationary Pareto exponent $\alpha_c$ as a function of the refresh ratio $\mu_r/J$ for the same two-state parameters as in Fig.~\ref{fig:tail_crossover}: $J=1$, $\Dlow=0.12$, $\Dhigh=1.10$, and $\beta=0.72$.  The curve is obtained analytically as the smallest root above unity of $\Delta_\alpha=0$.  It connects the slow-refresh limit $\alpha_c^{\mathrm{slow}}=1+J/\Dhigh\simeq1.91$ to the fast-refresh limit $\alpha_c^{\mathrm{fast}}=1+J/\Dbar\simeq3.54$, where $\Dbar=\beta\Dlow+(1-\beta)\Dhigh=0.3944$.}
    \label{fig:qc_vs_refresh}
\end{figure}

For any finite $\mu_r > 0$, the stationary tail remains a single algebraic mode of the coupled two-channel system. Eq.~\eqref{eq:q-slow} is the limit of this exponent as $\mu_r \to 0$. Figure~\ref{fig:tail_crossover} shows the stationary complementary cumulative distribution for several refresh rates, and Fig.~\ref{fig:qc_vs_refresh} shows the exact tail exponent $\alpha_c$ obtained from $\Delta_\alpha=0$ as a function of the refresh ratio. The exponent satisfies $\alpha_c^{\mathrm{slow}}<\alpha_c<\alpha_c^{\mathrm{fast}}$ for every finite $\mu_r>0$, as we check directly from the determinant. At $\alpha=1$, $F_\ell(1)=F_h(1)=J$, so
\begin{equation}
\Delta_1=J^2+\mu_r J>0.
\end{equation}
For $1<\alpha<\alpha_c^{\mathrm{slow}}$, the three factors $F_\ell(\alpha)$, $F_h(\alpha)$, and $\alpha J-\alpha(\alpha-1)\Dbar$ are all positive, so $\Delta_\alpha>0$ and no root exists in this interval. At $\alpha=\alpha_c^{\mathrm{slow}}$, $F_h$ vanishes by definition, and the remaining term gives
\begin{equation}
\Delta_{\alpha_c^{\mathrm{slow}}}=\mu_r\,\alpha_c^{\mathrm{slow}}\,J\left(1-\frac{\Dbar}{\Dhigh}\right)>0.
\end{equation}
At $\alpha=\alpha_c^{\mathrm{fast}}$, the refresh term vanishes by definition, and
\begin{equation}
\Delta_{\alpha_c^{\mathrm{fast}}}=F_\ell(\alpha_c^{\mathrm{fast}})\,F_h(\alpha_c^{\mathrm{fast}})<0,
\end{equation}
since $\Dlow<\Dbar<\Dhigh$ gives $F_\ell>0$ and $F_h<0$ at this point. By continuity, the first root above unity satisfies
\begin{equation}
\alpha_c^{\mathrm{slow}}<\alpha_c<\alpha_c^{\mathrm{fast}}
\end{equation}
for every finite $\mu_r>0$. Since $\alpha_c<\alpha_c^{\mathrm{fast}}<1+J/\Dlow$, we have $F_\ell(\alpha_c)>0$, and the first row of the system gives $a_h/a_\ell=[F_\ell(\alpha_c)+\mu_r(1-\beta)]/(\mu_r\beta)>0$, so the tail mode has positive weights.

\section{Conclusion}
\label{sec:conclusion}

We have shown that switching diffusivity produces a transient non-Gaussian shape for isolated multiplicative growth, while its persistence can remain relevant to stationary tails once redistribution is introduced. At the single-agent level, the logarithmic process subordinates to the integrated diffusivity $\Lambda_t = \int_0^t D_s\,\dd s$ and reduces, for an arbitrary redraw law, to a Gaussian mixture in the diffusing-diffusivity / DSCTRW class \cite{arutkin2024doubly, chechkin2017brownian, chubynsky2014diffusing,sposini2018random}. The process therefore exhibits a non-Gaussian short-time crossover, while the long-time distribution of the logarithmic process self-averages in shape and becomes Gaussian, with an effective variance still controlled by the fluctuations and persistence of $D_t$. At the many-agent level, this self-averaging no longer controls the stationary tail selection. For an arbitrary redraw law, the stationary large-$x$ problem leads to a spectral condition for admissible algebraic modes. For the two-state model, the physical Pareto tail can be selected explicitly: its exponent is the first root above unity of $\Delta_\alpha$ and interpolates between the fast-refresh Bouchaud--M\'ezard value $1+J/\bar D$ \cite{bouchaud2000wealth} and the slow-refresh value $1+J/\Dhigh$, remaining strictly below the former at every finite refresh rate. The complete stationary distributions in both asymptotic regimes further show that averaging the diffusivity before constructing the stationary state is valid only in the fast-refresh limit.

Several extensions are natural. The mean-field reduction relies on the independence of the $D_i$ across agents; a common-mode volatility factor, relevant in financial markets and in macroeconomic fluctuations with aggregate shocks, breaks this independence. The all-to-all connectivity assumption can be relaxed toward sparse exchange networks, where the condensation regime of the Bouchaud--M\'ezard model ~\cite{bouchaud2000wealth, ichinomiya2013power, medo2009breakdown, hur2024interplay,burda2002wealth,garlaschelli2008effects} must be revisited in the presence of latent volatility; a directed polymer mapping in the spirit of Derrida and Spohn offers a natural point of entry~\cite{derrida1988polymers}. Finally, the transient relaxation spectrum, first-passage statistics, and large deviations of the coupled system are accessible through the same matrix structure that yielded $\Delta_\alpha$ and deserve separate treatment, in parallel with recent results on switching diffusions~\cite{gueneau2025large}.

Beyond these theoretical extensions, the model suggests a possible empirical implication. The classical Bouchaud--M\'ezard value $1+J/\bar D$, recovered here only in the fast-refresh limit, has served as a reference exponent for Pareto tails of individual wealth \cite{bouchaud2023self,klass2007forbes,benhabib2011distribution} and, via Zipf-like variants, of firm sizes \cite{axtell2001zipf, gabaix2009power}. Empirical multiplicative systems are known to exhibit persistent and heterogeneous volatility, with firm growth rates displaying size-dependent variance scaling $\sigma(S) \sim S^{-\gamma}$ and financial returns displaying long-range volatility correlations \cite{bouchaud2001power,gabaix2016dynamics}. Within the present model, persistence in the diffusivity provides a stationary mechanism by which systems with the same mean diffusivity $\bar D$ can exhibit different Pareto exponents. Calibrating the two-state parameters $(\Dlow, \Dhigh, \beta, \mu_r)$ against candidate datasets on firm sizes or financial wealth, and testing whether observed departures from the Bouchaud--M\'ezard prediction are accounted for by switching diffusivity, is left for future work.

\begin{acknowledgments}
M.A. acknowledges discussions with Lazlo Dubois-Jacquemin on wealth redistribution.
This work was supported by the National Science and Technology Council, the Ministry of Education (Higher Education Sprout Project NTU-115L104022-1), and the National Center for Theoretical Sciences of Taiwan.
\end{acknowledgments}

\bibliography{bibli}

\end{document}

\title{Supplemental Material for ``Switching diffusivity selects Pareto tail exponent in random growth with redistribution''}
\author{Maxence Arutkin}
\affiliation{School of Chemistry, Tel Aviv University, Tel Aviv, Israel}
\affiliation{Physics Division, National Center for Theoretical Sciences, National Taiwan University, Taipei 106319, Taiwan}
\author{Alexandre Vall\'ee}
\affiliation{Department of Epidemiology and Public Health, Foch Hospital, Suresnes, France}
\maketitle
\tableofcontents
\section{Exponential redraw law}
\label{sec:SM-exponential}

We now specialize the general renewal result derived in the main text to the exponential redraw law $f_D(D)=D_0^{-1}e^{-D/D_0}$ for $D\geq0$.

\subsection{Explicit Fourier--Laplace propagator}

For the exponential law, averaging the single-plateau factor over the diffusivity gives $\int_0^\infty f_D(D)e^{-(k^2+ik)D\tau}\,\mathrm{d}D=[1+D_0(k^2+ik)\tau]^{-1}$. The remaining Laplace integral can be evaluated in terms of the exponential integral $E_1(z)=\int_z^\infty u^{-1}e^{-u}\,\mathrm{d}u$, yielding
\begin{equation}
\widetilde{\phi}_{\mathrm{exp}}(k,s)=\frac{1}{D_0(k^2+ik)}\exp\!\left[\frac{s+\mu_r-ikm}{D_0(k^2+ik)}\right]E_1\!\left[\frac{s+\mu_r-ikm}{D_0(k^2+ik)}\right].
\label{eq:SM-exp-phi}
\end{equation}
Substitution into the general renewal expression gives the exact Fourier--Laplace propagator
\begin{equation}
\widetilde{\Psi}_{\mathrm{exp}}(k,s)=\frac{\widetilde{\phi}_{\mathrm{exp}}(k,s)}{1-\mu_r\widetilde{\phi}_{\mathrm{exp}}(k,s)}.
\label{eq:SM-exp-propagator}
\end{equation}

\subsection{Short-time distribution}

For $t\ll\mu_r^{-1}$, the diffusivity is approximately frozen at its initial draw. Defining the centered log-increment as $r=y_t-y_0-mt$, its distribution is the Gaussian mixture
\begin{equation}
p_{\mathrm{short}}^{\mathrm{exp}}(r,t)=\int_0^\infty\frac{e^{-D/D_0}}{D_0\sqrt{4\pi Dt}}\exp\!\left[-\frac{(r+Dt)^2}{4Dt}\right]\mathrm{d}D.
\label{eq:SM-exp-short-mixture}
\end{equation}
The frozen-diffusivity approximation is a standard Gaussian-mixture construction~\cite{chechkin2017brownian,cherstvy2016anomalous}. For the exponential redraw law, evaluating this mixture gives
\begin{equation}
p_{\mathrm{short}}^{\mathrm{exp}}(r,t)=\frac{e^{-r/2}}{2D_0\sqrt{t(D_0^{-1}+t/4)}}\exp\!\left[-|r|\sqrt{\frac{1}{D_0t}+\frac{1}{4}}\right].
\label{eq:SM-exp-short}
\end{equation}
The resulting asymmetric Laplace form is the frozen-diffusivity limit of the exponential mixture. Its asymmetry originates from the It\^o drift contribution $-D_t\,\mathrm{d}t$ in the logarithmic dynamics.

\subsection{Long-time Gaussian limit}

For the exponential law, $\bar D=D_0$ and $\sigma_D^2=D_0^2$. Since each Poisson event redraws the diffusivity independently, $\mathrm{Cov}(D_s,D_{s+u})=D_0^2e^{-\mu_r u}$. The variance of the integrated diffusivity is therefore
\begin{equation}
\mathrm{Var}(\Lambda_t)=2D_0^2\left[\frac{t}{\mu_r}-\frac{1-e^{-\mu_r t}}{\mu_r^2}\right].
\label{eq:SM-exp-var-Lambda}
\end{equation}
At long times, $\mathrm{Var}(\Lambda_t)\simeq2D_0^2t/\mu_r$. The Brownian contribution has variance $2\mathbb{E}[\Lambda_t]=2D_0t$, while its covariance with $\Lambda_t$ vanishes. Consequently,
\begin{equation}
\frac{y_t-y_0-(m-D_0)t}{\sqrt{t}}\sim\mathcal{N}\!\left(0,\,2D_0+\frac{2D_0^2}{\mu_r}\right).
\label{eq:SM-exp-long}
\end{equation}
Thus the logarithmic process becomes Gaussian at long times, while its asymptotic variance retains the contribution of persistent diffusivity fluctuations.

\subsection{Finite-time divergence of annealed moments}

Conditional on the diffusivity trajectory, the $n$th moment is $\mathbb{E}[z_t^n\mid\{D_s\}]=z_0^n\exp[nmt+n(n-1)\Lambda_t]$. For $n>1$, this quantity grows exponentially with the integrated diffusivity. Restricting the average to trajectories with no refresh in $[0,t]$, which occur with probability $e^{-\mu_r t}$ and satisfy $\Lambda_t=Dt$, gives
\begin{equation}
\mathbb{E}[z_t^n]\geq z_0^n e^{nmt-\mu_r t}\int_0^\infty\frac{e^{-D/D_0}}{D_0}e^{n(n-1)Dt}\,\mathrm{d}D.
\label{eq:SM-exp-moment-bound}
\end{equation}
The integral diverges when $n(n-1)D_0t\geq1$. Hence every annealed moment of order $n>1$ becomes infinite no later than
\begin{equation}
t_n^\ast=\frac{1}{n(n-1)D_0}.
\label{eq:SM-exp-moment-time}
\end{equation}
This finite-time divergence results from the unbounded exponential tail of the redraw law rather than from the renewal mechanism itself. It motivates the bounded diffusivity distribution used for the interacting stationary problem.
\section{General stationary formulation for arbitrary diffusivity laws}
\label{sec:SM-general-stationary}

We consider the relative-wealth dynamics
\begin{equation}
\mathrm{d}x=J(1-x)\,\mathrm{d}t+\sqrt{2D_t}\,x\,\mathrm{d}B_t.
\end{equation}
The diffusivity $D_t$ is renewed at Poisson rate $\mu_r$, with each new value drawn independently from a normalized distribution $f_D(D)$. We denote by $p(x,D,t)$ the joint density of relative wealth and diffusivity, and by $P(x,t)=\int_0^\infty p(x,D,t)\,\mathrm{d}D$ its marginal over $D$.

\subsection{Joint evolution equation}

Between two renewal events, $D$ remains constant. At fixed diffusivity, the dynamics in $x$ is governed by the drift $J(1-x)$ and the multiplicative diffusion coefficient $Dx^2$. The renewal process removes probability from the current diffusivity state at rate $\mu_r p(x,D,t)$. Conversely, trajectories renewed from any previous diffusivity state are assigned the value $D$ with density $f_D(D)$, giving an incoming rate $\mu_r f_D(D)P(x,t)$.

The joint forward equation is therefore
\begin{equation}
\partial_t p(x,D,t)=-\partial_x\!\left[J(1-x)p(x,D,t)\right]+D\,\partial_x^2\!\left[x^2p(x,D,t)\right]+\mu_r\!\left[f_D(D)P(x,t)-p(x,D,t)\right].
\label{eq:SM-general-joint}
\end{equation}
At stationarity,
\begin{equation}
0=-\partial_x\!\left[J(1-x)p(x,D)\right]+D\,\partial_x^2\!\left[x^2p(x,D)\right]+\mu_r\!\left[f_D(D)P(x)-p(x,D)\right].
\label{eq:SM-general-stationary}
\end{equation}

\subsection{Algebraic modes of the large-$x$ operator}

For large $x$, the constant part of the restoring drift is subleading, so the stationary equation contains terms with the same scaling in $x$. We therefore seek a separable algebraic mode
\begin{equation}
p_\alpha(x,D)=a_\alpha(D)x^{-1-\alpha}.
\label{eq:SM-general-mode}
\end{equation}
If $a_\alpha(D)$ is integrable, the associated marginal contribution is $P_\alpha(x)=A_\alpha x^{-1-\alpha}$, with $A_\alpha=\int_0^\infty a_\alpha(D)\,\mathrm{d}D$.

For this mode,
\begin{equation}
\partial_x\!\left[xp_\alpha(x,D)\right]=-\alpha a_\alpha(D)x^{-1-\alpha},
\end{equation}
and
\begin{equation}
\partial_x^2\!\left[x^2p_\alpha(x,D)\right]=\alpha(\alpha-1)a_\alpha(D)x^{-1-\alpha}.
\end{equation}
Keeping the leading large-$x$ terms in Eq.~\eqref{eq:SM-general-stationary} gives
\begin{equation}
\left[\mu_r+\alpha J-\alpha(\alpha-1)D\right]a_\alpha(D)=\mu_r f_D(D)A_\alpha.
\label{eq:SM-general-amplitude}
\end{equation}
Hence
\begin{equation}
a_\alpha(D)=\frac{\mu_r f_D(D)}{\mu_r+\alpha J-\alpha(\alpha-1)D}A_\alpha.
\label{eq:SM-general-amplitude-solution}
\end{equation}
Integrating over $D$ and using the definition of $A_\alpha$ gives the spectral condition
\begin{equation}
1=\mu_r\int_0^\infty\frac{f_D(D)}{\mu_r+\alpha J-\alpha(\alpha-1)D}\,\mathrm{d}D.
\label{eq:SM-general-tail-condition}
\end{equation}

\subsection{Admissibility of the general mode}

A solution of Eq.~\eqref{eq:SM-general-tail-condition} does not automatically correspond to a physical stationary tail. The associated amplitude $a_\alpha(D)$ must be non-negative and integrable over the support of $f_D$. In particular, a candidate root is inadmissible if the denominator in Eq.~\eqref{eq:SM-general-amplitude-solution} vanishes inside the relevant support or produces a sign-changing or non-integrable mode.

Equation~\eqref{eq:SM-general-tail-condition} should therefore be interpreted as a spectral condition for admissible algebraic modes of the asymptotic stationary operator. Establishing that one of these modes is realized as the tail of the full stationary distribution additionally requires existence of the stationary state and selection of the corresponding positive mode. We do not establish these properties here for an arbitrary redraw law.

\subsection{Reduction to the two-state model}

For the two-state redraw law
\begin{equation}
f_D(D)=\beta\delta(D-D_\ell)+(1-\beta)\delta(D-D_h),
\end{equation}
the general spectral condition becomes
\begin{equation}
1=\mu_r\left[\frac{\beta}{\mu_r+F_\ell(\alpha)}+\frac{1-\beta}{\mu_r+F_h(\alpha)}\right],
\label{eq:SM-general-binary}
\end{equation}
where $F_i(\alpha)=\alpha J-\alpha(\alpha-1)D_i$. Multiplying by the two denominators gives
\begin{equation}
F_\ell(\alpha)F_h(\alpha)+\mu_r\left[\beta F_\ell(\alpha)+(1-\beta)F_h(\alpha)\right]=0.
\end{equation}
Using $\bar D=\beta D_\ell+(1-\beta)D_h$, this reduces to
\begin{equation}
F_\ell(\alpha)F_h(\alpha)+\mu_r\left[\alpha J-\alpha(\alpha-1)\bar D\right]=0.
\label{eq:SM-general-binary-determinant}
\end{equation}
This is exactly the determinant condition obtained from the explicit two-channel formulation in the main text. The positivity of the corresponding two-component mode and the location of the physical root are established there. The limiting stationary distributions and their finite-refresh corrections are derived in the following section.

\section{Asymptotic stationary regimes}
The stationary distributions can be obtained analytically in the limits of slow and fast diffusivity refresh. We denote by
\begin{equation}
    p_D(x)= \frac{(J/D)^{1+J/D}}{\Gamma(1+J/D)} x^{-2-J/D} e^{-\frac{J}{Dx}}
\end{equation}
the normalized stationary solution for a constant diffusivity $D$, and define $\bar D=\beta D_\ell+(1-\beta)D_h$. 
\subsection{Slow-refresh expansion of the Pareto exponent}
For $\mu_r\ll J$, the diffusivity remains approximately constant over the redistribution time scale $J^{-1}$. Integrating the stationary channel equations shows that, for any $\mu_r>0$, $\int_0^\infty P_\ell(x)\,dx=\beta$ and $\int_0^\infty P_h(x)\,dx=1-\beta$. Taking the stationary limit first and then $\mu_r\to0$, the two channel densities approach their fixed-diffusivity stationary states, $P_\ell(x)\sim\beta p_{D_\ell}(x)$ and $P_h(x)\sim(1-\beta)p_{D_h}(x)$. The full stationary distribution in the slow-refresh limit is therefore $P_{\rm slow}(x)=\beta p_{D_\ell}(x)+(1-\beta)p_{D_h}(x)$, or explicitly
\begin{equation}
    P_{\rm slow}(x)=\beta\frac{(J/D_\ell)^{1+J/D_\ell}}{\Gamma(1+J/D_\ell)}x^{-2-J/D_\ell}e^{-\frac{J}{D_\ell x}}+(1-\beta)\frac{(J/D_h)^{1+J/D_h}}{\Gamma(1+J/D_h)}x^{-2-J/D_h}e^{-\frac{J}{D_hx}}.
\end{equation}
Since $D_h>D_\ell$, the high-diffusivity contribution dominates the far tail,
\begin{equation}
    P_{\rm slow}(x)\sim(1-\beta)\frac{(J/D_h)^{1+J/D_h}}{\Gamma(1+J/D_h)}x^{-2-J/D_h},
\end{equation}
so that the limiting Pareto exponent is $\alpha_{\rm slow}=1+\frac{J}{D_h}$. At $\mu_r=0$, the determinant also admits the root $1+J/D_\ell$, associated with the faster-decaying low-diffusivity contribution. For small finite $\mu_r$, we therefore follow the root $1+J/D_h$, which controls the far tail. The exact determinant condition derived in the main text can be written, for $\alpha>1$, as
\begin{equation}
    \alpha\left[J-(\alpha-1)D_\ell\right]\left[J-(\alpha-1)D_h\right]+\mu_r\left[J-(\alpha-1)\bar D\right]=0.
\end{equation}
At $\mu_r=0$, the root controlling the stationary tail is $\alpha=1+J/D_h$. Treating $\alpha$ as a function of $\mu_r$ and differentiating the equation at this point gives
\begin{equation}
    J\left(1-\frac{D_\ell}{D_h}\right)\left[-(J+D_h)\left.\frac{d\alpha}{d\mu_r}\right|_{\mu_r=0}+\beta\right]=0,
\end{equation}
Hence
$\left.\frac{d\alpha}{d\mu_r}\right|_{\mu_r=0}=\frac{\beta}{J+D_h},$
and
$\alpha\simeq1+\frac{J}{D_h}+\frac{\beta\mu_r}{J+D_h}$
when $\mu_r\ll J$. 
\subsection{Fast-refresh expansion by adiabatic elimination}
For $\mu_r\gg J$, the diffusivity state relaxes rapidly compared with the wealth distribution. We define $P=P_\ell+P_h$ and $Q=P_\ell-\beta P$, so that $P_\ell=\beta P+Q$ and $P_h=(1-\beta)P-Q$. With $\mathcal L_i f=-\partial_x[J(1-x)f]+D_i\partial_x^2[x^2f]$, the stationary channel equations give
\begin{equation}
0=[\beta\mathcal L_\ell+(1-\beta)\mathcal L_h]P+(\mathcal L_\ell-\mathcal L_h)Q,
\label{eq:SM-fast-P}
\end{equation}
and
\begin{equation}
0=\beta(1-\beta)(\mathcal L_\ell-\mathcal L_h)P+[(1-\beta)\mathcal L_\ell+\beta\mathcal L_h]Q-\mu_rQ.
\label{eq:SM-fast-Q}
\end{equation}
As $\mu_r\to\infty$ at fixed $J$, $D_\ell$, $D_h$, and $\beta$, the refresh term $-\mu_rQ$ balances the source $\beta(1-\beta)(\mathcal L_\ell-\mathcal L_h)P$ to leading order, giving
\begin{equation}
Q\simeq\frac{\beta(1-\beta)}{\mu_r}(\mathcal L_\ell-\mathcal L_h)P=-\frac{\beta(1-\beta)(D_h-D_\ell)}{\mu_r}\partial_x^2(x^2P).
\end{equation}
Thus $Q\to0$ as $\mu_r\to\infty$, and the channel densities approach $P_\ell=\beta P$ and $P_h=(1-\beta)P$. The marginal equation then reduces to the Bouchaud--M\'ezard equation with diffusivity $\bar D$. Its stationary zero-current condition, $\bar D\,\partial_x(x^2P)=J(1-x)P$, yields the normalized distribution
\begin{equation}
P_{\rm fast}(x)=p_{\bar D}(x)=\frac{(J/\bar D)^{1+J/\bar D}}{\Gamma(1+J/\bar D)}x^{-2-J/\bar D}e^{-J/(\bar D x)}.
\end{equation}
This is the complete stationary distribution in the fast-refresh limit, with Pareto exponent $\alpha_{\rm fast}=1+J/\bar D$.

The first finite-refresh correction to the Pareto exponent follows from the exact determinant condition derived in the main text. Dividing this condition by $\mu_r$ gives
\begin{equation}
J-(\alpha-1)\bar D+\frac{\alpha}{\mu_r}[J-(\alpha-1)D_\ell][J-(\alpha-1)D_h]=0.
\end{equation}
Writing $\varepsilon=\mu_r^{-1}$, the fast-refresh limit corresponds to $\varepsilon=0$, where $\alpha=\alpha_{\rm fast}=1+J/\bar D$. Differentiating the condition with respect to $\varepsilon$ at this point yields
\begin{equation}
\left.\frac{d\alpha}{d\varepsilon}\right|_{\varepsilon=0}=\frac{\alpha_{\rm fast}}{\bar D}\left[J-\frac{J}{\bar D}D_\ell\right]\left[J-\frac{J}{\bar D}D_h\right].
\end{equation}
Using $\bar D-D_\ell=(1-\beta)(D_h-D_\ell)$ and $\bar D-D_h=-\beta(D_h-D_\ell)$, we obtain
\begin{equation}
\alpha\simeq1+\frac{J}{\bar D}-\frac{\beta(1-\beta)(D_h-D_\ell)^2}{\mu_r\bar D}\left(\frac{J}{\bar D}\right)^2\left(1+\frac{J}{\bar D}\right).
\end{equation}
The correction is negative: at large but finite refresh rate, the stationary Pareto tail decays more slowly than in the strict fast-refresh limit.

The same expression for $Q$ determines the relative contribution of the diffusivity states in the far tail:
\begin{equation}
\lim_{x\to\infty}\frac{Q(x)}{P(x)}\simeq-\frac{\beta(1-\beta)(D_h-D_\ell)}{\mu_r}\left(\frac{J}{\bar D}\right)\left(1+\frac{J}{\bar D}\right).
\end{equation}
Because $P_\ell/P=\beta+Q/P$ and $P_h/P=1-\beta-Q/P$, the high-diffusivity state is overrepresented among large-wealth events, although its population weight remains $1-\beta$. The inverse-gamma density describes the strict fast-refresh limit. At large but finite refresh rate, the far tail carries the corrected exponent above. 

\section{Sampling procedures}
\label{sec:SM-sampling}

\subsection{Direct sampling at fixed time for Fig.~1}

Figure~1 is obtained by sampling directly the distribution at a prescribed time $t$, without discretizing the dynamics. The number $N$ of diffusivity renewals in $[0,t]$ is first drawn from a Poisson distribution of mean $\mu_r t$. Conditional on $N$, the $N+1$ plateau durations are uniformly distributed over the simplex. Each plateau is independently assigned $D_\ell$ with probability $\beta$ or $D_h$ with probability $1-\beta$.

If $M$ plateaux are in the high-diffusivity state, then $U=0$ for $M=0$, $U=1$ for $M=N+1$, and otherwise the fraction $U$ of the total interval spent in that state follows a beta distribution with parameters $M$ and $N+1-M$. The integrated diffusivity is therefore sampled exactly as
\begin{equation}
\Lambda_t=t\left[D_\ell+(D_h-D_\ell)U\right].
\label{eq:SM-sampling-Lambda}
\end{equation}

Conditional on $\Lambda_t$, the variable $r_t$ is Gaussian and is sampled as
\begin{equation}
r_t=-\Lambda_t+\sqrt{2\Lambda_t}\,Z,
\label{eq:SM-sampling-r}
\end{equation}
where $Z$ is a standard Gaussian variable. This procedure samples the one-time distribution exactly. For each value of $\mu_r t$, $10^7$ independent realizations are generated. 

\subsection{Stationary sampling for Fig.~2}

For Fig.~2, we simulate the relative wealth $x$ using a symmetric splitting of the redistribution and multiplicative-noise dynamics. The redistribution equation $\dot x=J(1-x)$ has the exact flow
\begin{equation}
R_a(x)=1+(x-1)e^{-Ja}.
\label{eq:SM-redistribution-flow}
\end{equation}
Over one interval $dt$, conditional on the integrated diffusivity $\Lambda_n=\int_{t_n}^{t_n+dt}D_s\,ds$, the multiplicative part is sampled exactly as
\begin{equation}
M_n(x)=x\exp\!\left[-\Lambda_n+\sqrt{2\Lambda_n}\,Z_n\right],
\label{eq:SM-multiplicative-flow}
\end{equation}
where $Z_n$ is a standard Gaussian variable. One numerical step is then
\begin{equation}
x_{n+1}=R_{dt/2}\!\left(M_n\!\left(R_{dt/2}(x_n)\right)\right).
\label{eq:SM-numerical-update}
\end{equation}

The integrated diffusivity $\Lambda_n$ is sampled exactly within each interval. We first draw the number $n_r$ of diffusivity renewals from a Poisson distribution of mean $\mu_r dt$. If $n_r=0$, the diffusivity remains unchanged throughout the interval. If $n_r\geq1$, the interval contains $n_r+1$ plateaux: the first keeps the diffusivity inherited from the preceding interval, while after each renewal a new value is drawn independently, with probabilities $\beta$ and $1-\beta$ for $D_\ell$ and $D_h$, respectively. Conditional on $n_r$, the relative plateau durations are uniformly distributed over the simplex. If $K$ plateaux are in the high-diffusivity state, their total occupation fraction $U$ therefore follows a beta distribution with parameters $K$ and $n_r+1-K$. The integrated diffusivity is then $\Lambda_n=dt[D_\ell+(D_h-D_\ell)U]$, with $U=0$ or $1$ when all plateaux are respectively in the low or high-diffusivity state. The diffusivity drawn at the last renewal is retained for the next interval. Thus the renewal process within $dt$ is sampled exactly, including the possibility of multiple renewals; the remaining time-discretization error arises only from the splitting between redistribution and multiplicative evolution.

Each realization starts from $x=1$, while the initial diffusivity is drawn from its stationary two-state distribution. We evolve each realization for $T_{\rm eq}=10/J$, corresponding to ten redistribution times; the mean dependence on the initial value of $x$ is then reduced by a factor $e^{-10}$. We use $dt=0.025/\max(J,D_h)$ for the symmetric splitting scheme.

For each refresh rate, $2\times10^6$ independent realizations are generated and a single final value is retained from each realization. All values are used in the empirical complementary cumulative distribution, with no tail cutoff. The smallest empirical probability is therefore $5\times10^{-7}$, while a probability of $10^{-5}$ contains on average twenty realizations. The values $\mu_r/J=0.1,3,12,$ and $100$ are used to show one curve close to each asymptotic regime and two clearly separated intermediate cases.

\bibliography{bibli}